\documentclass[]{interact}

\usepackage{epstopdf}

\usepackage[numbers,sort&compress]{natbib}
\bibpunct[, ]{[}{]}{,}{n}{,}{,}

\usepackage{graphicx,stfloats}
\usepackage{color}
\usepackage{lineno,hyperref}
\usepackage{amssymb}
\usepackage{amsmath}
\usepackage{mathtools}
\usepackage{bm}
\usepackage{xcolor}

\begin{document}

\articletype{}

\title{Efficient Simulations of Propagating Flames Using Adaptive Mesh Refinement}

\author{
\name{Caelan Lapointe \textsuperscript{a}\thanks{CONTACT Caelan Lapointe. Email: caelan.lapointe@colorado.edu}, Nicholas~T.~Wimer \textsuperscript{a}, Jeffrey~F.~Glusman \textsuperscript{a}, Prakriti Sardana \textsuperscript{a}, Amanda~S.~Makowiecki \textsuperscript{a}, John~W.~Daily \textsuperscript{a},  Gregory~B.~Rieker \textsuperscript{a}, Peter~E.~Hamlington \textsuperscript{a}}
\affil{\textsuperscript{a} Paul M.~Rady Department of Mechanical Engineering, University of Colorado, Boulder, CO 80309, USA}
}

\maketitle

\begin{abstract}
Wildland fires are complex multi-physics problems that span wide spatial scale ranges. Capturing this complexity in computationally affordable numerical simulations for process studies and ``outer-loop'' techniques (e.g., optimization and uncertainty quantification) is a fundamental challenge in reacting flow research. Further complications arise for propagating fires where \textit{a priori} knowledge of the fire spread rate and direction is typically not available. In such cases, static mesh refinement at all possible fire locations is a computationally inefficient approach to bridging the wide range of spatial scales relevant to wildland fire behavior. In the present study, we address this challenge by incorporating adaptive mesh refinement (AMR) in fireFoam, an OpenFOAM solver for simulations of complex fire phenomena. The AMR functionality in the extended solver, called wildFireFoam, allows us to dynamically track regions of interest and to avoid inefficient over-resolution of areas far from a propagating flame. We demonstrate the AMR capability for fire spread on vertical panels and for large-scale fire propagation on a variable-slope surface that is representative of real topography. We show that the AMR solver reproduces results obtained using much larger statically refined meshes, at a substantially reduced computational cost.
\end{abstract}


\section{Introduction}
\label{Introduction}
Wildland fires have become increasingly common and impactful to human life and property, motivating a renewed focus on the detailed study of fire dynamics \cite{Brown2018} to support suppression and mitigation efforts. Given the potential hazard and cost of experimental studies, particularly at large spatial scales, computational simulations are an attractive approach to studying fire behavior. However, computationally efficient full-physics simulations that avoid empirical \textit{ad hoc} models \cite{hanson_potential_2000} are required to study fire dynamics over widely varying and realistic conditions, as well as to enable the reliable use of ``outer-loop'' techniques such as optimization (e.g., for mitigation and suppression efforts) and uncertainty quantification (e.g., for risk assessment).

Perhaps the greatest difficulty in performing such high-fidelity simulations is the wide spatio-temporal scale separation present in wildland fire problems. Disparate temporal scales, from relatively slow flow advection to fast chemical reactions, are often addressed by assuming infinitely fast chemistry. This mixed-is-burnt assumption can be made because chemical reactions occur on much shorter timescales than the turbulent mixing timescale for non-premixed flames. The eddy dissipation model \cite{chen_extension_2014, chen_large_2014} can then be used to relate the combustion timescale to local turbulence timescales. Flamelet models or other mechanism reduction approaches (e.g., at run-time \cite{pope_computationally_1997,singer_operator-splitting_2006,singer_exploiting_2004} or \textit{a priori} \cite{niemeyer_skeletal_2010,glusman2019}) can also be used to balance fidelity with computational cost.

Relevant spatial scales range from the small-scale structures of solid fuels to the large scales characteristic of topographical variations and atmospheric dynamics. Methods to address temporal scale separation generally increase the minimum computational timestep, either by not resolving the reaction timescale or by making finite-rate chemistry more affordable. However, the minimum spatial length scale is often dictated by geometric constraints, such as the fuel size or local geography. Although static mesh refinement can be used to address some of this scale separation, such approaches are inadequate in fire spread problems where the region of interest may vary rapidly and is not known \textit{a priori}. 

In the present study, we use adaptive mesh refinement (AMR) to address the challenge of bridging wide spatial scale separations in simulations of fire spread. Using AMR, computational resources (and, by extension, grid cells and physical resolution) can be concentrated where they are most needed, for example to track a propagating flame. Here we specifically focus on fire spread over pyrolyzing surfaces and introduce a new extension of the OpenFOAM solver fireFoam \cite{wang_large_2011,repository_firefoam_2018}, called wildFireFoam. We demonstrate this AMR-enabled solver, which is related to the AMR solver diffusionFireFoam \cite{lapointe2020efficient}, for meter-scale vertical fire spread on combustible panels, as well as for fire spread over variable-slope terrain at scales from tens to hundreds of meters. We show, in particular, that the solver reproduces results from static mesh simulations with the same fine-scale resolution, but at a substantially reduced computational cost. The new solver is available to the fire modeling community at the public GitHub repository \url{https://github.com/clapointe2011/public} for simulations of a broad range of fire spread problems. 

\section{Computational framework: wildFireFoam}
OpenFOAM is an open-source, finite-volume computational fluid dynamics platform that includes both pre- and post-processing functionalities, in addition to a wide array of solvers \cite{repository_openfoam_2018}. In the present study, we implement AMR in fireFoam, a solver developed by FM Global to simulate complex fire phenomena \cite{wang_large_2011,repository_firefoam_2018}. Designed for simulations of industrial fire problems, fireFoam has been used previously to study problems ranging from non-reacting buoyant plumes \cite{maragkos_application_2012,maragkos2013large} and pool fires with static \cite{wang_large_2011,maragkos_large_2017,maragkos2017implementation} and dynamic meshes (i.e., using AMR) \cite{lapointe2020efficient}, to larger room-scale fires \cite{vilfayeau_numerical_2015,salmon2018firefoam}, pyrolysis modeling \cite{ding_modeling_2015,ding_large_2014,meredith_comprehensive_2013,fukumoto2018large,trouve_large_2010}, and fire suppression \cite{ren_firefoam_nodate,meredith_comprehensive_2013}.

\subsection{Physical treatment}
FireFoam is typically used to perform large eddy simulations (LES), and this is also the approach that we take with the AMR-enabled solver. In the gas phase, three-dimensional Favre-filtered compressible Navier-Stokes equations and conservation equations for mass, total enthalpy, and species (with or without reactions) are solved on a single computational mesh. These equations are given as
\begin{align}
\frac{\partial \rho}{\partial t} + \frac{\partial}{\partial x_i}(\rho u_i) &= 0,\label{gas:continuity} \\
\frac{\partial (\rho u_i)}{\partial t} + \frac{\partial}{\partial x_j}(\rho u_i u_j) &= -\frac{\partial p_rgh}{\partial x_i} + \frac{\partial \tau_{ij}}{\partial x_j} + \rho g_i\,, \label{gas:momentum}\\
\frac{\partial (\rho h +\rho K)}{\partial t} + \frac{\partial}{\partial x_i}(\rho u_i h +\rho u_i K) &= \frac{\partial p}{\partial t} + \frac{\partial}{\partial x_i}\left[\alpha_\mathrm{eff}\frac{\partial (\rho h)}{\partial x_i}\right] + Q_\mathrm{rxn} + Q_\mathrm{rad}\,, \label{gas:enthalpy} \\
\frac{\partial (\rho Y_i)}{\partial t} + \frac{\partial}{\partial x_j}(\rho Y_i u_j) &= \frac{\partial}{\partial x_j} \left[ \alpha_\mathrm{eff} \frac{\partial (\rho Y_i)}{\partial x_j}\right]+ \rho \omega_i\,, \label{gas:species}
\end{align}
where $\tau_{ij}$ is the viscous and turbulent stress tensor given by
\begin{equation}
    \tau_{ij} = \mu_\mathrm{eff} \left(\frac{\partial u_i}{\partial x_j} + \frac{\partial u_j}{\partial x_i}\right) - \frac{2}{3} \mu_\mathrm{eff} \frac{\partial u_k}{\partial x_k} \delta_{ij}\,,
    \label{gas:tauNS}
\end{equation}
and the Lewis number is assumed to be unity. In the above equations, $\rho$ is the density, $u_i$ is the velocity, $p_\mathrm{rgh}$ is the dynamic pressure, $p$ is the pressure (i.e., $p=p_\mathrm{rgh}+\rho g_i x_i+p_\mathrm{ref}$, where $p_\mathrm{ref}$ is a reference pressure), $g_i$ is the gravitational acceleration, $h$ is the sensible enthalpy per unit volume, $K$ is the kinetic energy, $\alpha_\mathrm{eff}$ and $\mu_\mathrm{eff}$ are the turbulent thermal diffusivity and viscosity, respectively, $Q_\mathrm{rxn}$ and $Q_\mathrm{rad}$ represent heat addition due to reactions and radiation, respectively, $Y_i$ is the mass fraction of the $i^\mathrm{th}$ species, and $\omega_i$ is the reaction rate for the $i^\mathrm{th}$ species.

Depending on the choice of turbulence closure used to compute $\mu_\mathrm{eff}$ and $\alpha_\mathrm{eff}$, sub-grid scale (SGS) transport equations may also be necessary to close the governing equations. As in the standard version of fireFoam, we use mixture-averaged density and transport properties; the latter are temperature-dependent and computed using empirical correlations. Combustion modeling is assumed to be infinitely fast for non-premixed problems and governed by an eddy dissipation model. Radiation modeling is accomplished using a finite volume implementation of the discrete ordinate method and a grey mean model for absorption and emission, commonly with a prescribed radiative fraction. For a detailed description of the gas phase physical treatment in fireFoam, we refer the interested reader to \cite{wang_large_2011}. Details of the solution process for gas-phase simulations with AMR are provided in \cite{lapointe2020efficient}.

In the solid phase, one-dimensional conservation equations are solved for mass, enthalpy, and species, given as \cite{chaos_evaluation_2011,fukumoto2018large,vinayak_mathematical_2017}
\begin{align}
\frac{\partial \rho_s}{\partial t} &= -R_g\,, \label{solid:continuity}\\
\frac{\partial (\rho_s h_s)}{\partial t}  - \frac{\partial}{\partial x}\left[ \frac{\partial (\kappa_s T_s)}{\partial x} \right] &= S_\mathrm{rad} + S_\mathrm{rxn} + S_\mathrm{gas} +  S_\mathrm{flux} \label{solid:energy}\\
\frac{\partial (\rho_s Y_i)}{\partial t} &= R_i\,. \label{solid:species}
\end{align}
Here, $\rho_s$ and $h_s$ are solid-phase density and specific sensible enthalpy, $Y_i$ is the mass fraction of solid species $i$, $R_i$ is the reaction rate of the $i^\mathrm{th}$ solid species, $R_g$ is the corresponding rate of production of pyrolysate, and $\kappa_s$ is the solid-phase thermal conductivity. Solid-phase energy source terms include an optional in-depth radiative source, $S_\mathrm{rad}$, a reaction source term, $S_\mathrm{rxn}$, a source term to account for mass loss (or, equivalently, gas production) from the solid phase, $S_\mathrm{gas}$, and an optional enthalpy flux-based source term for gas motion within the solid, $S_\mathrm{flux}$. These equations are solved on a series of independent, one-dimensional solid meshes using mixture-averaged properties. That is, gas flux through the solid is treated implicitly, and is computed from the solid mass loss rate (or equivalently from pyrolysate production rate) \cite{fukumoto2018large, vinayak_mathematical_2017,chaos_evaluation_2011}.

The solid phase is coupled to the gas phase via mapped boundary conditions for the velocity, temperature, and fuel. For velocity, the mass flux of pyrolysate in the solid region is matched to a corresponding mass flux of fuel in the gaseous region using either a one-to-one mapping \cite{repository_openfoam_2018} or an energetically-equivalent mapping \cite{repository_firefoam_2018}. For temperature, heat transfer is coupled between the solid and gaseous regions, and for the fuel, pyrolysate in the solid region is converted to a known fuel in the gaseous region. Additional details of the solid phase modeling are available in Refs.~\cite{fukumoto2018large,vinayak_mathematical_2017,chaos_evaluation_2011}.

\subsection{Numerical treatment\label{subsec:num}}
In the gas phase, the conservation equations are solved in a segregated manner and coupled using a procedure (i.e., PIMPLE) involving outer SIMPLE iterations and inner PISO correctors \cite{wang_large_2011}. Both the temporal integration and the spatial discretization have second-order accuracy. Total variation diminishing (TVD) schemes for the divergence of both bounded and unbounded conserved scalar quantities, as well as a stabilized central difference scheme for the velocity convective term, are employed to aid stability. Additionally, in the interest of stability, enthalpy gradients are limited. Diffusive terms are unbounded and second-order with limited non-orthogonal corrections to aid stability.

For the solid phase, we again use second-order methods for temporal integration and spatial discretization. Gas-solid coupling is treated in a segregated manner; solid phase conservation equations are solved first, followed by solution of the gas phase. A global timestep is computed from a user-specified maximum advective Courant number in the gaseous region and maximum diffusion number in the solid region.

\subsection{Adaptive mesh refinement (AMR)\label{subsec:contribute}}
Previously, we developed an AMR solver for efficient simulations of turbulent diffusion flames, called diffusionFireFoam \cite{lapointe2020efficient}. In the present work, we have extended the diffusionFireFoam solver to combine the OpenFOAM dynamic meshing functionality described in Ref.~\cite{h._jasak_automatic_2000} with the full suite of physical models (e.g., pyrolysis, Lagrangian particle, and surface film modeling) included in fireFoam. To use AMR with mapped boundary conditions (necessary to couple solid and gaseous regions), faces on any coupled boundary (and, by extension, the cells containing said faces) are protected from refinement; they must be pre-refined to the highest level of AMR that will be used. We have additionally included functionality to constrain refinement to specific regions of the domain; this is particularly useful for the panel spread cases included in this work where refinement may be constrained to be near the solid panels.

The AMR implementation in OpenFOAM takes the form of a single, unstructured mesh for solution of the gas phase (a static mesh is used for the solid phase, given the low cost of one-dimensional computations). At user-specified intervals, the mesh is updated based on a single refinement field. Values of the field between lower and upper bounds are marked for refinement. Similarly, values of the field below a cutoff are identified for coarsening. Other input parameters include the refinement frequency and a parameter to delay two-to-one refinement (i.e., ``buffer'' refinement around added cells). We refer to Refs.~\cite{h._jasak_automatic_2000} and \cite{lapointe2020efficient} for a more complete description of the AMR process, including verification and validation of the approach in a range of both non-reacting and reacting flow problems.

Previously, we extended the AMR library in Ref.~\cite{h._jasak_automatic_2000} to enable refinement and coarsening on an arbitrary number of fields \cite{lapointe2020efficient}. These fields are scaled at run-time to range between $0$ and $1$; refinement bounds are therefore percents of the maximum value (e.g., the top $99.9\%$ of values at a given time, or lower and upper bounds of $0.001$ and $1$, respectively). Prior work also involved extensive testing of the AMR procedure, focusing on flux corrections for new cell faces to ensure mass conservation. All of these improvements have also been implemented in wildFireFoam, and in the subsequent sections we describe the specific setup of the AMR procedure for each test case.

\section{Verification: Single panel vertical fire spread}
In order for the present implementation of AMR in wildFireFoam to be deemed successful, it must result in a reduction of simulation cost with little or no impact on accuracy, as compared to equivalently-refined static mesh simulations.

We first compare the cost and accuracy of AMR in wildFireFoam against a statically-refined simulation of a vertical fire spread problem based on a tutorial included with the standard distribution of OpenFOAM \cite{repository_openfoam_2018}. This case places a $60$~kW propane burner in the center of a narrow initially quiescent domain. The domain is bounded on two sides by combustible panels (composed of homogenous mixtures of unburnt, solid fuel and burnt, charred product) and is open on the remaining sides. The propane burns infinitely fast as it mixes with ambient air, heating the wood. The wood pyrolyzes and combustible products are transferred to the gaseous region, where they burn and the flame spreads. 

\subsection{Statically refined simulation}
\begin{figure}[t!]
\centering
\includegraphics[width=0.7\textwidth]{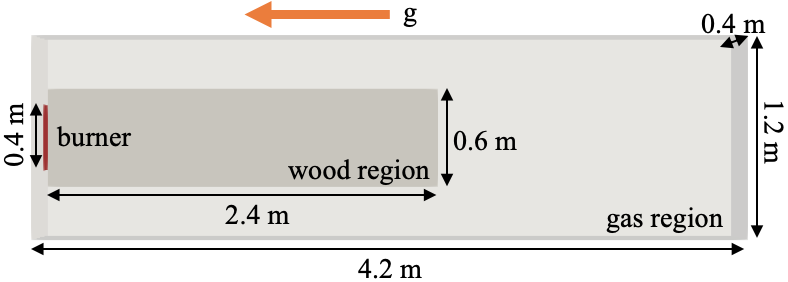}
\caption{Schematic of the vertical fire spread case, showing the propane burner in red, as well as both the wood and gas regions. The orange arrow indicates the direction of gravity. \label{fig:1}}
\end{figure}

\begin{figure}[t!]
\centering
\includegraphics[width=0.8\textwidth]{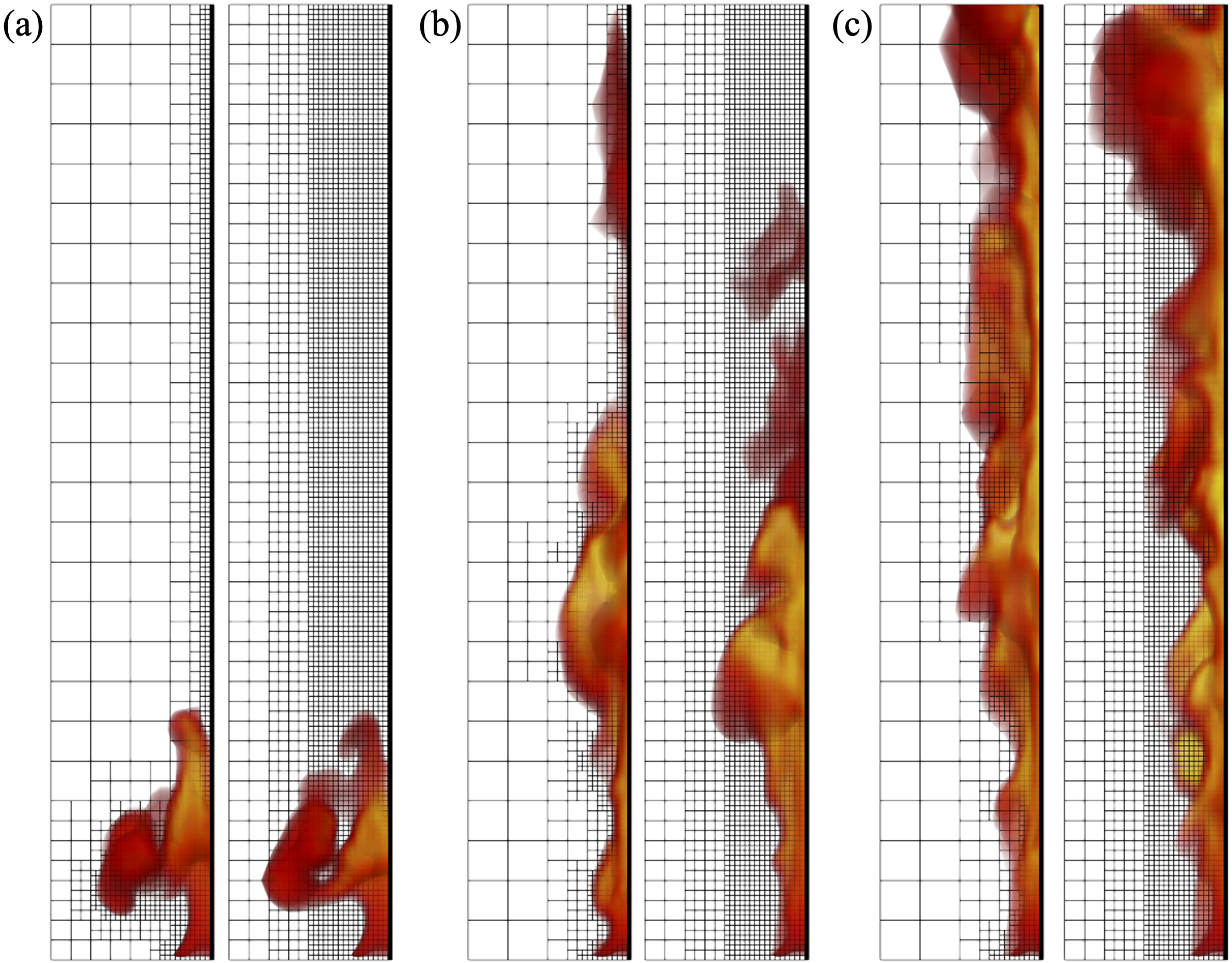}
\caption{Instantaneous temperature volume renders for AMR (left sub-panels) and static (right sub-panel) vertical panel spread cases at times (a) 0.9~s, (b) 2.5~s, and (c) 7.5~s. Temperature is colored from low (black) to high (yellow). The computational meshes are visualized in a plane through the center of the domain in each panel. \label{fig:vertSpreadViz}}
\end{figure}

For the present demonstration of wildFireFoam, we modify the tutorial case and focus on a single panel for simplicity, as shown in Fig.~\ref{fig:1}. A base mesh measuring $[0.4\times 4.2\times 1.2]$~m, centered on the propane burner, was created for the solution of the gaseous region with $10$~cm base resolution. Three nested, static regions of refinement were then created, including a highest refinement region with $1.25$-cm resolution occupying a $[0.2\times 2.4\times 0.6]$~m volume. Successive refinements occupied $[0.3\times 0.3\times 0.8]$~m and $[0.4\times 3.6\times 1.0]$~m volumes with $2.5$~cm and $5.0$~cm resolutions, respectively. Figure \ref{fig:vertSpreadViz} shows slices through the center of the resulting static mesh at $z=0$, indicating the successive regions of static refinement in the $x-y$ plane.

The solid panel region was created using native OpenFOAM mesh extrusion utilities based on a collection of faces on the boundary of the base mesh, within the $1.25$~cm resolution region. A $[0.6\times2.4]$~m area was extruded by $1$~cm, with $1$~mm resolution in the third dimension. Next, the propane burner was created in a $[0.1\times0.4]$~m area on the bottom boundary. 

In total, the static mesh for the gaseous region was composed of $181,440$ cells and the solid region was composed of $92,160$ cells. We ran the simulation in parallel on $3$ processors for $15$~s of physical time.

Gas and solid regions were coupled as described in Section \ref{subsec:num}. The solid-gas interface is treated as a wall and we used wall functions for SGS viscosity and thermal diffusivity from \cite{repository_openfoam_2018}. Within the gas phase, turbulence modeling was accomplished using a one equation eddy-viscosity model \cite{yoshizawa_statistical_1986} with $C_k$ of $0.03$ \cite{ren2017large}. Combustion was modeled using the eddy dissipation method, as implemented in \cite{repository_firefoam_2018}, with coefficients $C_\mathrm{EDC}$, $C_\mathrm{Diff}$, and $C_\mathrm{Stiff}$ of $4.0$, $0.4$, and $10^{-10}$, respectively \cite{vilfayeau2015large,ren_large_2016}. Radiation modeling was performed using the finite volume discrete ordinate method with a prescribed radiative fraction of $0.2$. In the solid phase, a single-step, finite-rate Arrhenius reaction was used according to the reaction 
\[\mathrm{wood}^{4.86}\rightarrow \mathrm{char}+\mathrm{gas}\,,\] 
where the rate is given as a function of temperature $T$ by $7.83\times 10^{10}\exp(-15274.57/T)$. Pyrolysis is assumed to only occur once the solid has reached the critical temperature of $400$~K; this pyrolysis model is identical to that found in the baseline tutorial case. Temporal and spatial discretization is nominally second-order in the gas phase, as described in Section \ref{subsec:num}. Three outer PIMPLE and two inner PISO iterations are used to solve the gas-phase conservation equations \cite{wang_large_2011}. The solid phase is set numerically (with the exception of temporal integration, which is changed to be second-order) following the example cases in \cite{repository_firefoam_2018}. A maximum convective Courant number of $0.4$ is enforced in the gas phase, and a maximum diffusion number of $0.25$ is used for the solid phase; the most limiting constraint is used to compute a single global timestep. 


\subsection{AMR simulation\label{subsec:amr1}}
The physical setup of the AMR case is identical to that for the static case, described in the previous section. For the AMR, a base mesh of $10$~cm is created and only the boundary faces along the gas-solid interface are refined three levels to $1.25$~cm resolution, for an initial cell count of $24,360$. The solid regions are therefore identical between the static mesh and AMR simulations. 

Within the AMR simulation, three levels of refinement are used in the gaseous region to achieve $1.25$~cm resolution, refining on gas-phase heat release rate (HRR) as a measure of combustion strength and location (top $99.5\%$ of values), gradients in HRR (top $99.5\%$ of values) and enstrophy (top $90\%$ of values). These refinement thresholds are based on those determined for AMR simulations of pool fires described in Ref.~\cite{lapointe2020efficient}, where it was found that the AMR simulations were in good agreement with both static mesh simulations as well as experimental results. 

The mesh is updated every three flow iterations, and one buffer layer is used. Refinement fields are scaled at run-time to more easily allow assignment of lower and upper refinement bounds. Refinement is also constrained to be near the panel to prevent unnecessary cell addition. The AMR case is run in serial on a single processor, also for $15$~s, with identical physical and numerical settings as the static simulation; tighter (i.e., lower) tolerances are employed for linear solvers to ensure convergence for a given mesh. It should be noted that other refinement settings (i.e., fields and refinement criteria) may give even better results and these values should be taken as illustrative rather than optimal. 

\subsection{Results and discussion}
Instantaneous 3D renderings of the flame spread up the panel are shown in Fig.~\ref{fig:vertSpreadViz} for the AMR and static simulations at $0.9$, $2.5$, and $7.5$~s after ignition. In general, the flame structure is qualitatively similar between the two simulations, with the propagation of the flame up the panel occurring at roughly the same rate in both cases. 

The resulting AMR meshes at three successive times in the simulations are also shown in Fig.~\ref{fig:vertSpreadViz}, illustrating the dynamically varying nature of the mesh and the gradually increasing number of grid cells in the AMR case as the simulation proceeds. The AMR case started with approximately $24,000$ cells in the gaseous region and, once spun up, fluctuated around $60,000$ cells, requiring $5.6$ cpu hours per physical second of simulation time, on average. Consequently, we were able to decrease the computational cost relative to the static simulation by roughly $200\%$, where the static case required an average of $16.27$~cpu-hours per second of simulation time.  

\begin{figure}[t!]
\centering
\includegraphics[scale=1]{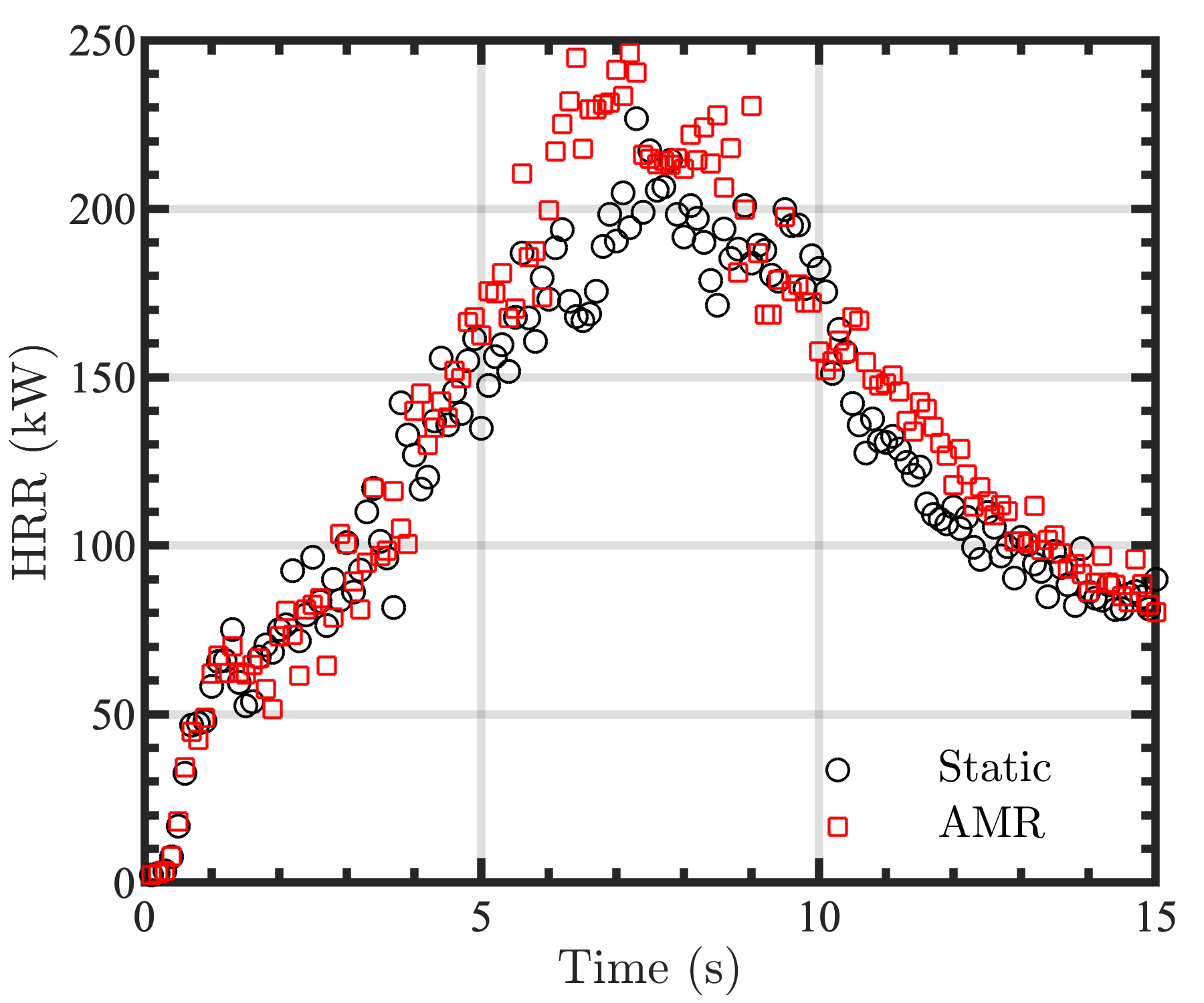}
\caption{Time history of volume-integrated heat release rate (HRR) for static and AMR simulations for the single-panel vertical fire spread case. \label{fig:3_2}}
\end{figure}

From a quantitative perspective, the time history of cumulative gas-phase HRR is compared for the static and AMR simulations in Fig.~\ref{fig:3_2}. There is generally good agreement between the two simulations, despite a small over-prediction of peak HRR in the AMR simulation. It is emphasized, however, that the flame spread simulations are turbulent and some variability in the exact HRR time series is expected, independent of the mesh treatment. The overall shape of the time series are very similar between the two cases, and there is close agreement in the final values of HRR after 15~s.

\begin{figure}[t!]
\centering
\includegraphics[scale=1]{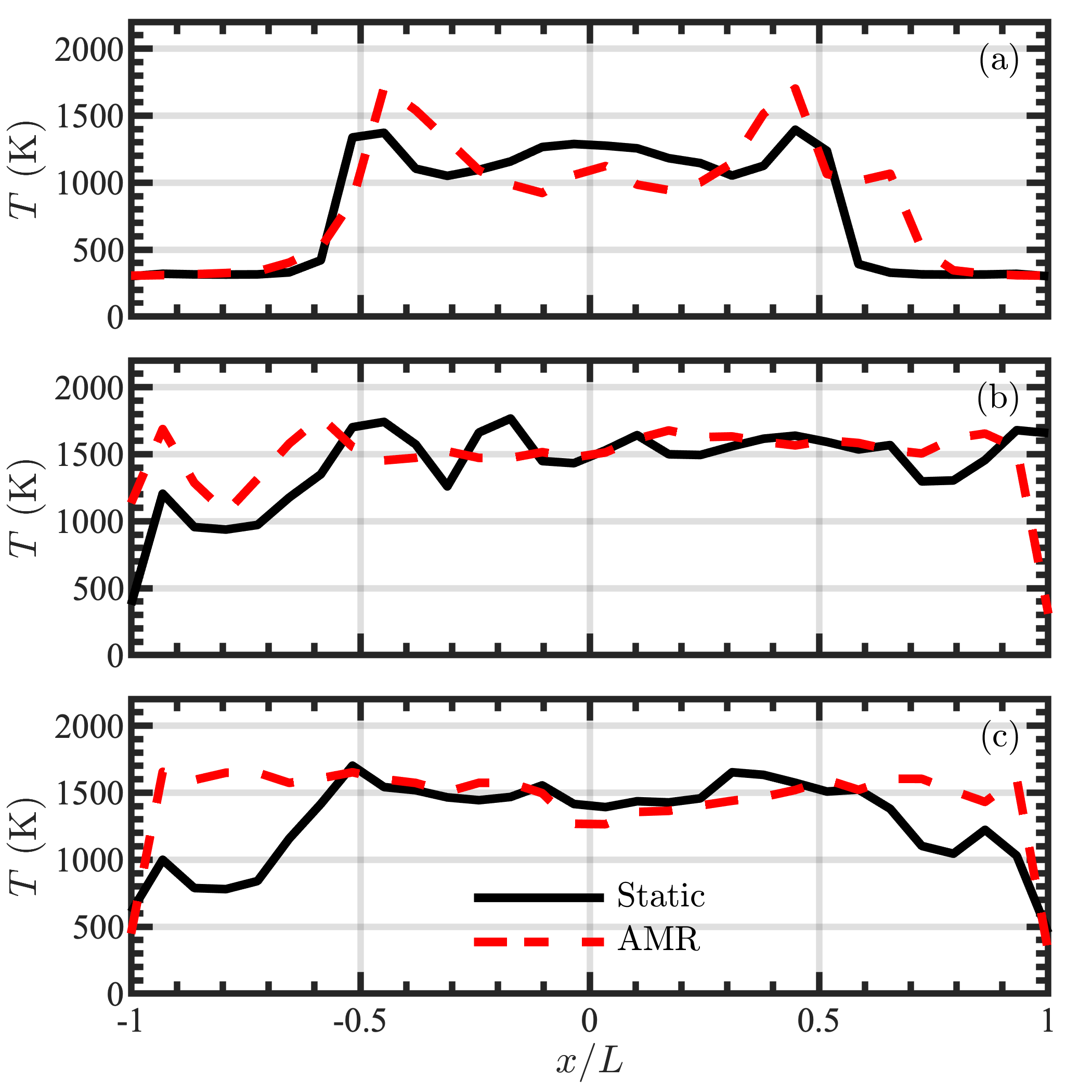}
\caption{Instantaneous gas-phase temperature near the gas-solid interface at $10$~s. Panels from top to bottom show profiles at panel heights of $z/H=0.1$, $0.5$, and $0.9$, where $H=2.4$~m is the total height. The transverse location $x$ is normalized by the half panel width $L=0.3$~m. \label{fig:4}}
\end{figure}

Finally, Fig.~\ref{fig:4} shows instantaneous temperature profiles from the AMR and static simulations at $8$~s (when HRR is close to its maximum) near the gas-solid interface at various heights above the burner. Once again, the temperatures are similar between the two simulations, particularly near the centerline (i.e., $x/L=0$), and deviations between the two sets of results are only observed near the edges of the flame. However, the turbulent nature of these simulations is again expected to create local and instantaneous differences in the simulation results, and the overall correspondence between the AMR and static simulations in Fig.~\ref{fig:4} suggests that the AMR simulation is accurately capturing the flame dynamics. 

It is important to note that the computational savings of the AMR simulation are dependent on the choice of refinement fields and their bounds. We have shown here that, for the choices outlined in Section \ref{subsec:amr1}, the AMR simulation provides results that are comparable to the static mesh results, at a $200\%$ lower computational cost. Additional improvements in efficiency may be achievable for other choices of refinement fields and AMR thresholds, although it should also be noted that further improvements in static mesh simulation cost may also be possible. As such, the present results should be taken as illustrative of the accuracy and efficiency of wildFireFoam, and greater (or lesser) gains may be realized in other problems. 

\subsection{Sensitivity to numerical and model choices}
Although wildFireFoam has been shown here to reproduce static mesh results at lower computational cost, the robustness of the AMR procedure to different numerical algorithms and model choices must also be considered in order to examine the broader utility of the solver. In the following, we consider, in particular, the effects of different spatial discretizations and different SGS models on the simulation results. Other sensitivities can also be considered, but are left as directions for future research. 

The spatial discretization of the vertical panel spread case was based on fireFoam settings in the OpenFOAM tutorials, namely a stabilized central differencing scheme for velocity divergence and TVD limited central differencing for scalar divergence quantities. These are now updated to be a filtered linear scheme for velocity and SGS turbulent kinetic energy (similar to \cite{vilfayeau2016large}) and the TVD Van Leer scheme \cite{van1974towards} for species, enthalpy, and resolved kinetic energy divergence calculations. Four levels of AMR are also used (with a corresponding decrease of one in the base level) to further increase computational savings. Given that the computational benefit of AMR has already been demonstrated, focus will again be placed on quantitative results.

\begin{figure}[!t]
\centering
\includegraphics[scale=1]{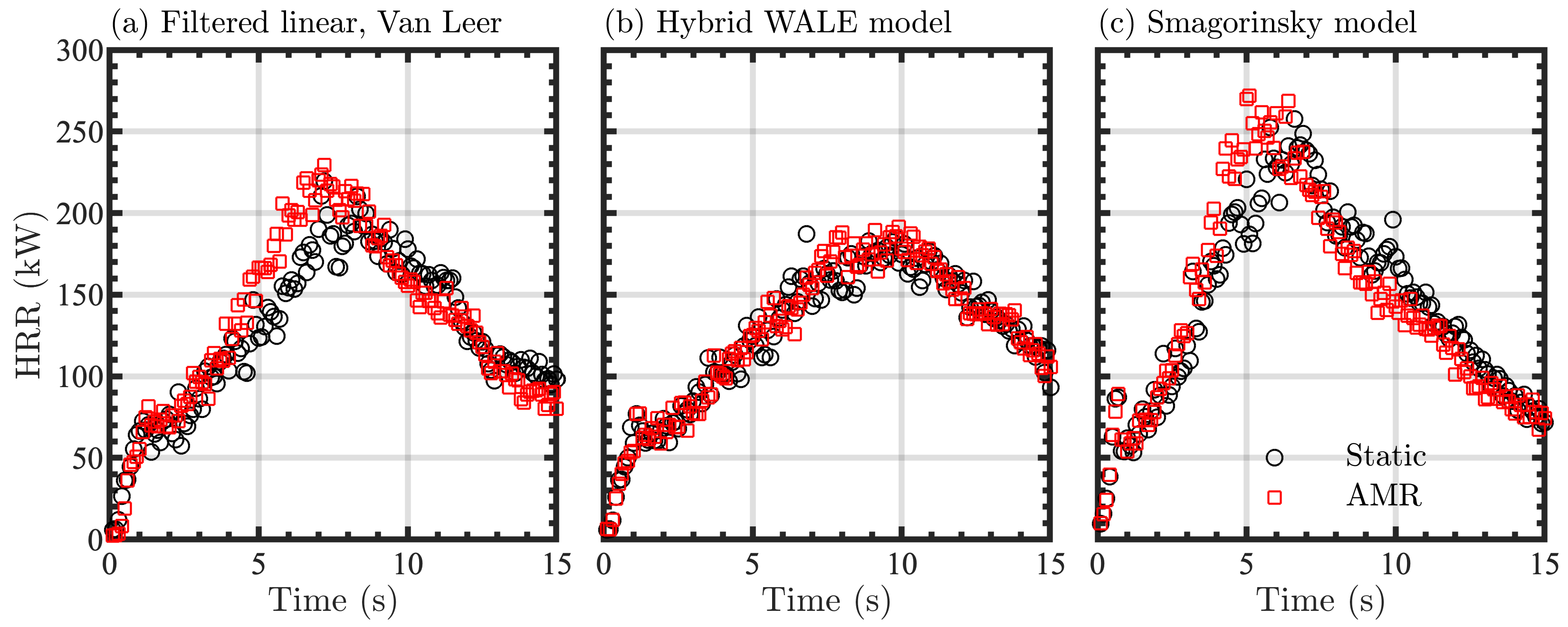}
\caption{Time history of volume-integrated heat release rate (HRR) for static and AMR simulations for the single-panel vertical fire spread case using (a) filtered linear and Van Leer schemes, (b) the hybrid WALE SGS model, and (c) the Smagorinsky SGS model. \label{fig:panelSpread_kEqn}}
\end{figure}

Results from this updated version of the verification case are presented in Fig.~\ref{fig:panelSpread_kEqn}(a), where integrated HRR for AMR and static simulations are shown, and fairly good agreement is again observed. The HRR in the AMR case does peak slightly before the peak HRR in the static case, although the general trends and magnitudes of the two time series remain similar. 

From the standpoint of physical modeling, the verification presented here was derived from fireFoam tutorials included with OpenFOAM, with little deviation in model choices. Given that other models (e.g., gas-phase turbulence modeling) may be desirable for other types of problems, it is pertinent to test them for use with wildFireFoam. Whereas most OpenFOAM tutorials employing an LES model rely on the one equation eddy viscosity model, the WALE model \cite{nicoud1999subgrid} has also been used for fireFoam simulations (e.g., \cite{ren_large_2016}). It is therefore worthwhile to show that the success of AMR fire spread cases such as those presented here are not impacted by turbulence model choice.

To test this sensitivity, we consider the hybrid WALE model (where SGS viscosity is computed algebraically but a transport equation for SGS turbulent kinetic energy is solved \cite{repository_firefoam_2018}), as well as the standard Smagorinsky model. The model-specific coefficients for the hybrid WALE model, $C_w$, and Smagorinsky model, $C_s$, are not changed from the OpenFOAM defaults of $0.325$ and $0.094$, respectively. The cases are otherwise identical to the updated single panel verification case presented above and AMR results will once again be compared to static results.

Integrated HRR for the hybrid WALE and Smagorinsky cases are shown in Figs.~\ref{fig:panelSpread_kEqn}(b) and (c), respectively. As with the baseline results in Fig.~\ref{fig:3_2}, there is overall good agreement between the static and AMR results, even as the HRR time series themselves change when using the different SGS models. To the latter point, the Smagorinsky results shown in Fig.~\ref{fig:panelSpread_kEqn}(c) exhibit a more rapid increase and decay in HRR than the other simulation restuls, including a larger maximum value fo HRR. Nevertheless, the choice of SGS model does not negatively impact the accuracy of AMR simulations as compared to static mesh simulations. The AMR implemented in wildFireFoam thus appears to be robust to different numerical and physical modeling choices.

\section{Verification: Large-scale wind-driven fire spread \label{sec:slope}}
For a second verification test of wildFireFoam, we demonstrate solver accuracy and computational savings for fire spread at larger scales over a variable-incline slope. With this case, we demonstrate the flexibility and utility of our AMR framework for efficient large-scale fire spread simulations, similar to \cite{canfield2014numerical} and recent studies using fireFoam to investigate fire-wind interactions~\cite{eftekharian2018cfd,eftekharian2019numerical,eftekharian2020correlations,eftekharian2020simulation}.

\subsection{Statically refined simulation}
For this verification case, we again performed both statically refined and AMR simulations. For the static case, we began with a $[147\times98\times98]$~m domain with uniform $0.875$~m resolution. We then deformed the bottom boundary to create a slope with a maximum incline of $27\%$, which is then extruded to create a solid region $1$~cm thick with $2$~mm resolution. The resulting gaseous and solid wood regions in the statically refined simulation are composed of $1,162,672$ and $93,910$ cells, respectively. 

An inflow profile for velocity characteristic of the atmospheric boundary layer was imposed to drive fire spread up the incline with $10$~m/s reference velocity and $10$~m reference height. The terrain surface was treated with a rough atmospheric wall function for the SGS eddy viscosity, and the flux of pyrolysate in the solid phase was directly transferred to propane in the gas phase. Otherwise, the solid-gas coupling was identical to the interface coupling for the first verification case, and the other vertical boundaries are open and the top is a slip wall. 

The gaseous and solid regions are also largely modeled identically to the first verification case, with changes to increase simplicity considering that exact, optimal model coefficients are unknown. Changes include using infinitely fast combustion modeling, using the default model coefficient for the one equation eddy viscosity model (i.e., $C_k = 0.094$), and the mapping of pyrolysate directly to gaseous fuel, as mentioned previously.

\begin{figure}[t!]
\centering
\includegraphics[width=0.6\textwidth]{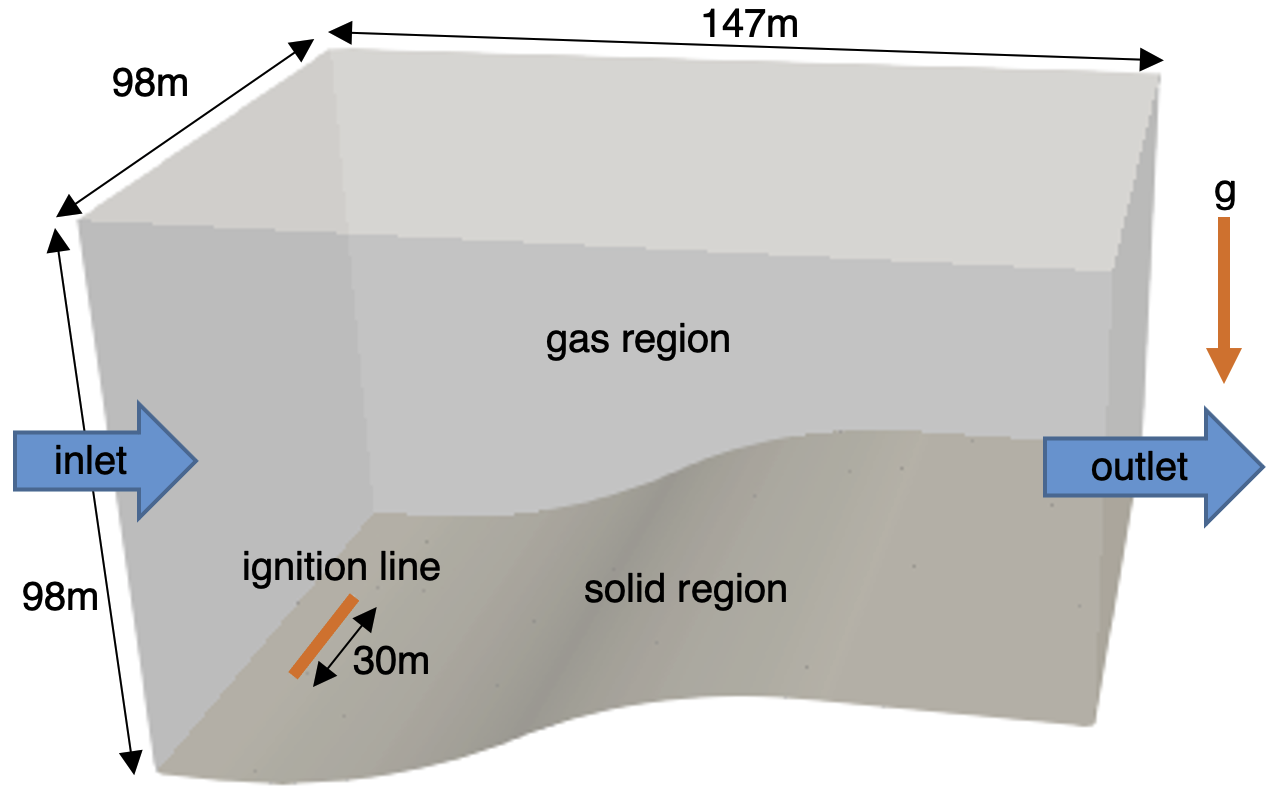}
\caption{Schematic of the large-scale fire spread case, showing the wind inlet and outlet as blue arrows, the ignition line in orange, and the direction of gravity by the orange arrow.} \label{fig:5}
\end{figure}

To trigger combustion, a $[30\times 1]$~m burner was created near the inlet to mimic a line fire with roughly $1.9$~MW/m source strength following \cite{eftekharian2019numerical}.  The simulation was run in parallel on $6$ processors for $20$~s of virtual time until the flame traveled up the incline and neared the outlet. A schematic of the computational domain, including predominant flow direction and preheated region, is shown in Fig.~\ref{fig:5}.

\subsection{AMR simulation}
For the AMR case, a coarse mesh with $7$~m resolution was created before refining the bottom solid boundary by three levels (giving $0.875$~m resolution) and deforming the boundary to produce the slope shown in Fig.~\ref{fig:5}. The numerical and physical setup in the solid region was identical to the static case, as is the line burner used for ignition. The gaseous region was initially composed of $45,962$ cells and three levels of refinement were used to provide $0.875$~m fine-scale resolution; the solid region is unchanged. As with the first verification case, we again refined on HRR (top $99.9\%$), its gradient (top $99.9\%$), and enstrophy (top $99\%$). The case was run on a single processor for $20$~s.

\subsection{Results and discussion}
\begin{figure}[t!]
\centering
\includegraphics[width=0.9\textwidth]{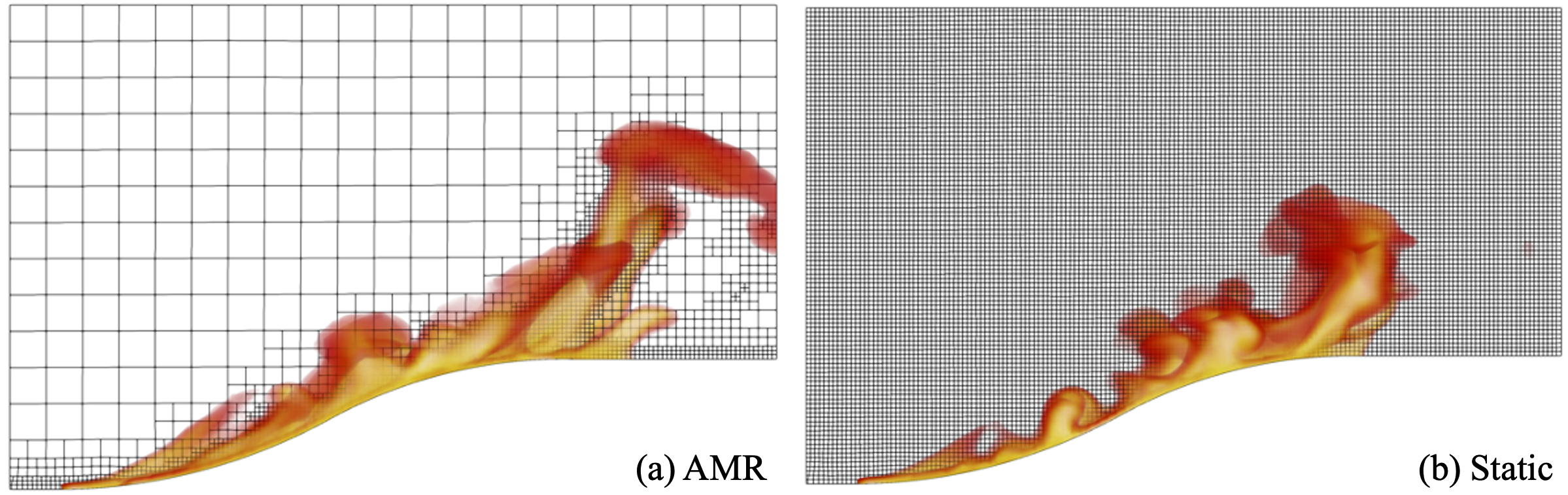}
\caption{Instantaneous temperature volume renders for AMR (a) and static (b) wind-driven fire spread cases. Temperature is colored from low (black) to high (yellow). The computational mesh is visualized for the center of the domain. \label{fig:terrainSpreadViz}}
\end{figure}

For this larger fire spread case, we first examine the two cases qualitatively. Temperature volume renders for AMR and static cases are shown in Fig.~\ref{fig:terrainSpreadViz} (a) and (b), respectively, with the computational mesh visualized in a plane through the center of the domain. Good qualitative agreement is again observed between the two cases, with the flame spreading at roughly equal rates along the terrain. 

\begin{figure}[!t]
\centering
\includegraphics[scale=1]{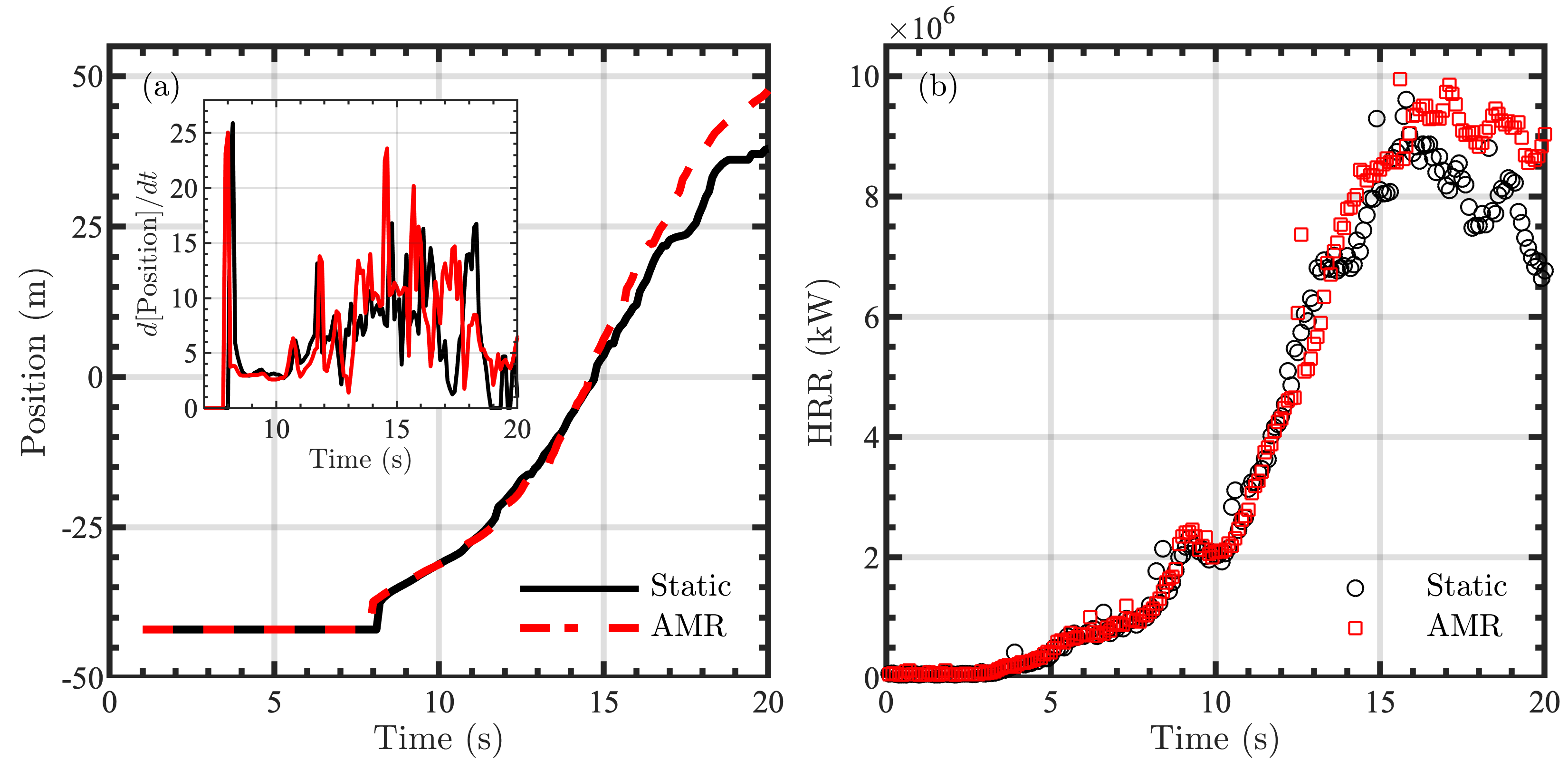}
\caption{(a) Time series of the flame front position on the slope of the large-scale fire spread case for the static and AMR simulations. The position is determined as the furthest down-wind extent of $0.01$ char values along the centerline. The inset shows the rate of spread versus time approximated with first-order finite differences. (b) Time series of volume-integrated HRR for the large-scale fire spread case for static and AMR simulations.  \label{fig:8}}
\end{figure}

This agreement can be demonstrated quantitatively by tracking the location of the flame front as it spreads up the slope. Computational data were post-processed to create high-resolution temperature profiles (with $1$~cm resolution) using pchip interpolation. Fire spread was assessed from char values above $0.01$ along the centerline of the gas-solid interface. The resulting time series of flame front position are compared in Fig.~\ref{fig:8}(a), indicating excellent agreement between the static and AMR cases. The inset of this figure also indicates that the rate of spread is comparable between the two cases. 

Similarly good agreement is observed in the time series of volume-integrated HRR in Fig.~\ref{fig:8}(b), where deviations between the AMR and static mesh results are only observed after the $15$~s mark as the flame front approaches the outlet. This small difference could be minimized in the future by moving the outlet further downstream or through the investigation of other outflow boundary conditions, including those that deactivate the AMR near the edge of the domain.

Figure~\ref{fig:terrainSpreadViz}(a) shows that the computational mesh for the AMR simulation closely tracks the hot gases and the flame spread. Whereas over $1,000,000$ cells and $11.42$ cpu hours per second of simulation time, on average, were required for the static case, the AMR case grew to roughly $140,000$ cells and only required an average of  approximately $0.92$ cpu hour per second of simulation time. The AMR case was therefore substantially cheaper. Once again, it should be noted that the mesh used for the static simulation could likely be tuned to decrease computational expense. However, we have shown that AMR can be used to accurately and efficiently capture fire spread with no \textit{a priori} knowledge of the flame location or evolution, and with no additional mesh tuning.

\section{Validation: Two-panel vertical fire spread}

Building on the previous single panel vertical spread verification case, we now validate wildFirefoam against experimental data for a panel spread case similar to the first verification case. In particular, a second panel with identical dimensions, opposite the first, was added to fully replicate the configuration in Ref.~\cite{chaos2011experiments}. For this configuration, two [$0.6\times2.4\times0.0154$]~m corrugated cardboard panels are separated by a [$0.3\times0.6$]~m $60$-kW propane burner. Experimental and computational data for chemical HRR and gas-solid interface temperature and heat flux were collected. Here we focus on HRR as a global metric of importance for combustion problems, and only perform an AMR simulation.

\subsection{AMR simulation}
The physical setup largely mimics the previous case shown in Fig.~\ref{fig:1} but now also includes a second solid region opposite the first. The gas-phase computational domain measures [$0.3\times4.8\times 2.4$]~m. Based on the recent success by Ren \emph{et al.}~\cite{ren2020convective} in simulating a similar case at coarse resolution, here we use a base resolution of $10$~cm and two levels of refinement to achieve an effective resolution of $2.5$~cm. 

As with the single panel verification case, the mesh boundaries adjacent to the panels on the gas-solid interface are pre-refined, for an initial cell count of $14,712$ cells, before meshing the solid region. The solid-phase computational domain is created as described previously, resulting in $4,608$ independent, one-dimensional regions, each $10$ cells in depth, for a total of $46,080$ cells.



The physical models (e.g., for SGS stresses and material properties) used in the gas phase are largely unchanged from the single panel spread verification case, and the radiative fraction was updated to be $0.22$ \cite{zeng2014radiation,ren2017large}. Solid phase thermophysical and chemistry properties are set following optimized values for the ``triple-wall`` corrugated cardboard reported in Ref.~\cite{chaos2011experiments}. The gas and solid phases are coupled as in the previous panel spread case with identical gas-solid interactions. 

Because the case needs to be run for a long physical time (the experiments extend to $150$~s \cite{chaos2011experiments}), numerical settings were adjusted to increase simulation speed, including increasing the maximum Courant number to $0.8$, using first-order Euler temporal integration in both phases, and using one outer PIMPLE iteration. The AMR settings were updated, as compared to the single-panel case, to use two levels of refinement with mesh updates every two flow iterations; refinement fields and bounds were, however, not changed from those used in the single-panel case. The AMR simulation was run in parallel on $3$ processors for $90$~s of physical time.

\subsection{Results and discussion}

\begin{figure}[t!]
\centering
\includegraphics[scale=1]{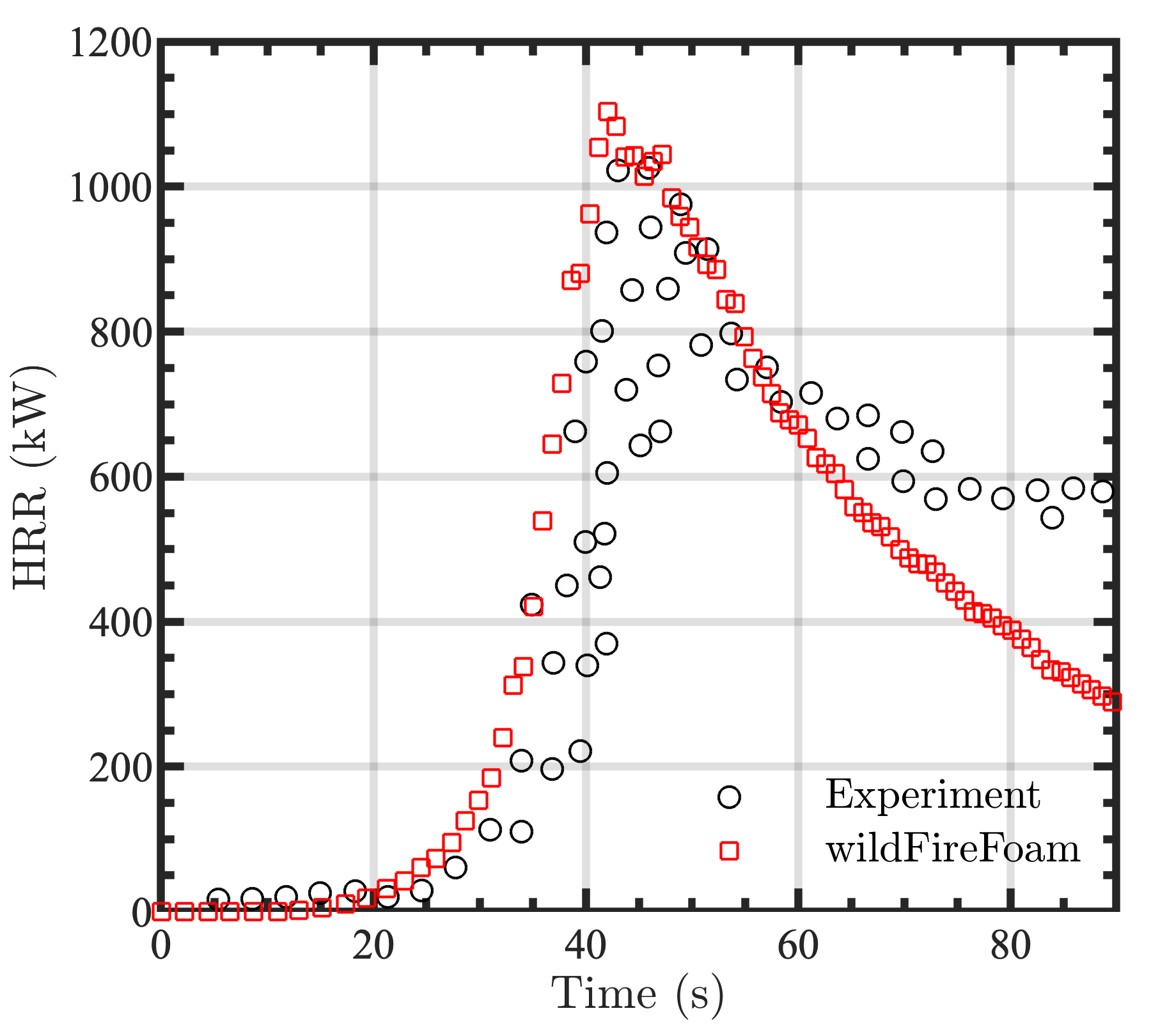}
\caption{Time history of chemical heat release rate (HRR) for the two-panel fire spread verification case. Experimental results \cite{chaos2011experiments} are shown as black circles and AMR simulation results from wildFireFoam are shown as red squares. \label{fig:9}}
\end{figure}

Experimental \cite{chaos2011experiments} and simulated chemical HRR (computed using the pyrolysate heat of combustion in \cite{ren2017large,wang2014numerical}) are compared in Fig.~\ref{fig:9}. After spinup of the simulation, a maximum cell count of roughly $30,000$ cells at peak HRR was required. Quantitatively, the AMR simulation provides a good prediction of the exponential increase in HRR (the primary goal of this simulation), matching computational results in \cite{chaos2011experiments}. The value of peak HRR is also similar between the experiments and simulations. Although a faster decay in HRR is observed in the simulations, as compared to the experiments, a similar behavior has also been seen in other simulations, specifically those using statically refined meshes \cite{chaos2011experiments}. As such, the difference in the decay rates is likely due to the physical modeling used in the simulation, rather than due to the use of AMR as opposed to static refinement.

In the future, this case could be improved, for example, by re-tuning solid-phase thermophysical parameters for use with wildFireFoam (and, by extension, a more recent version of OpenFOAM) to better match the experimental data after the time of peak heat release. In terms of modeling, possible improvements include the incorporation of a char oxidation model in the solid phase (e.g., as implemented in Ref.~\cite{repository_firefoam_2018}) and recent developments in wall models for convective heat transfer (e.g., \cite{ren2020convective}) and updated solid-phase thermophysical parameters (e.g., \cite{ren2017large,ren2020convective,wang2014numerical}). We anticipate that, given the results in Fig.~\ref{fig:9}, each of these improvements will improve the agreement between the simulation and experimental results, and that AMR can still be used to increase the computational cost of the simulations without negatively impacting accuracy.


\section{Conclusions and future work}
We have developed a novel extension of the OpenFOAM solver fireFoam that incorporates AMR and retains all physics originally included with the solver. We have demonstrated that this solver, called wildFireFoam, provides comparable agreement to static mesh simulations with the same fine-scale resolution for two verification cases and one validation case, at substantially reduced computational cost. Additionally, the vertical spread cases were based on existing OpenFOAM tutorials for fireFoam to showcase the easy addition of AMR to fireFoam simulations. Although we primarily highlight the computational benefits of using AMR in this work, we also note that it relaxes hardware requirements; the two dynamic verification cases were run on a single processor.

There are a number of ways the solver could be improved. To enable the use of AMR within the current OpenFOAM framework, mapped patches must be protected from refinement, and instead must be refined to the highest AMR level as a pre-processing step (prohibiting further refinement as part of the mesh update as an added precaution). This is necessary to couple the gas and solid regions and would require significant effort to allow the AMR to change the resolution of this patch (and therefore the resolution of the solid-phase). Another important point to note is that domain decomposition for allocation to computational cores must be done with care. In order to use the OpenFOAM utility to reconstruct an AMR mesh, cells with faces on a mapped patch must be placed on a single processor (or dynamic cases run in parallel may be mapped to a comparable uniform static mesh, as is done for the validation case in this work). This can lead to load imbalance, especially while the simulation is spinning up. Both of these points suggest, however, that further computational gains may be realized if load balancing is employed in tandem with AMR. Load-balancing extensions of the OpenFOAM AMR library have recently been published (see, e.g., \cite{rettenmaier2019load,blastfoam}) and will be incorporated into wildFireFoam in the future.

Future research directions include extensions of the simulations presented in this work, including detailed pyrolysis and soot modeling and larger parameter spaces (e.g., inflow velocities, initial fire-line sizes, solid fuel properties) for fire spread problems, as well as simulations incorporating Lagrangian particle and thin film modeling for the study of fire suppression. Additionally, ``outer-loop'' processes such as optimization and uncertainty quantification will now be feasible given the improvements in computational efficiency enabled by the use of AMR. Further investigation of linear solver tolerances as well as other divergence limiters and PIMPLE algorithm settings with respect to overall solver performance would also be beneficial.

\section*{Acknowledgements}
The authors would like to thank Yi Wang and Ning Ren of FM Global for their helpful input and for providing experimental data.

This work utilized the RMACC Summit supercomputer, which is supported by NSF (awards ACI-1532235 and ACI-1532236), the University of Colorado, Boulder, and Colorado State University.

\section*{Disclosure statement}
The authors declare no conflicts of interest.

\section*{Funding}
CL was supported by the NSF Graduate Research Fellowship Program under award DGE 1144083. NTW, JFG, ASM, JWD, GBR, and PEH were supported, in part, by SERDP under grant W912HQ-16-C-0026.

\end{document}